\documentstyle[12pt,epsfig]{article}
\newcommand{\be}{\begin{equation}}
\newcommand{\ee}{\end{equation}}
\begin{document}  
\topmargin 0pt
\oddsidemargin=-0.4truecm
\evensidemargin=-0.4truecm
\renewcommand{\thefootnote}{\fnsymbol{footnote}}
\newpage
\setcounter{page}{0}
\begin{titlepage}   
\vspace*{-2.0cm}  
%%%\vspace*{-1.0cm}
\begin{flushright}
%FT-UM-TH-03-02 \\
%FIS-UM-03-10 \\
%%\vspace*{-0.2cm}
hep-ph/0504069
\end{flushright}
%\vspace*{0.5cm}
\vspace*{0.1cm}
\begin{center}
{\Large \bf Low Energy Solar Neutrinos and Spin Flavour Precession
} \\ 
%\vspace{0.1cm}
%\vspace{0.16cm}
\vspace{0.6cm}

\vspace{0.4cm}

{\large 
Bhag C. Chauhan\footnote{On leave from Govt. Degree College, Karsog (H P) 
India 171304. E-mail: chauhan@cftp.ist.utl.pt},
Jo\~{a}o Pulido\footnote{E-mail: pulido@cftp.ist.utl.pt}\\
\vspace{0.15cm}
{  {\small \sl CENTRO DE F\'{I}SICA TE\'{O}RICA DAS PART\'{I}CULAS (CFTP) \\
 Departamento de F\'\i sica, Instituto Superior T\'ecnico \\
Av. Rovisco Pais, P-1049-001 Lisboa, Portugal}\\
}}
\vspace{0.25cm}
and \\
\vspace{0.25cm}
{\large R. S. Raghavan\footnote{E-mail: raghavan@vt.edu}} \\
{\it Department of Physics  \\ Virginia Polytechnic Institute and State University 
(Virginia Tech) \\  Blacksburg  VA 24060 USA}
\end{center}
\vglue 0.6truecm
\begin{abstract}
The possibility that the Gallium data effectively indicates a time modulation
of the solar active neutrino flux in possible connection to solar activity is
examined on the light of spin flavour precession to sterile neutrinos as a 
subdominant process in addition to oscillations. We distinguish two sets of 
Gallium data, relating them to high and low solar activity. Such modulation 
affects principally the low energy neutrinos ($pp$ and $^7 Be$) so that the 
effect, if it exists, will become most clear in the forthcoming Borexino and 
LENS experiments and will provide evidence for a neutrino magnetic moment. 
Using a model previously developed, we perform two separate fits in relation 
to low and high activity periods to all solar neutrino data. These fits 
include the very recent charged current spectrum from the SNO experiment. 
We also derive the model predictions for Borexino and LENS experiments.
\end{abstract}

\end{titlepage}   
\renewcommand{\thefootnote}{\arabic{footnote}}
\setcounter{footnote}{0}
\section{Introduction and Motivation} 

The quest for time dependence of the solar neutrino flux and  
the development of low energy ($<$ 1-2 MeV) solar neutrino experiments are
probably at present the major challenges facing solar neutrino physics. 
Evidence for time variability has been found by the Stanford Group 
\cite{Sturrock:2005wf},\cite{Sturrock:2004jv},\cite{Sturrock:2003kv}, 
\cite{Caldwell:2003dw}, \cite{Sturrock:2001qn} upon examination of time 
binned data from all experiments except, so far, SNO. If it is confirmed,
time variability will probably require neutrino spin flavor precession
(SFP) \cite{SFP} within the sun through the interaction of the neutrino 
magnetic moment with a varying solar magnetic field occuring in addition 
to the LMA effect. On the other hand the effort in real time experiments 
SuperKamiokande \cite{Fukuda:2002pe} and SNO \cite{Aharmim:2005gt} has been 
up to now concentrated 
in measuring the high energy $^8B$ flux which accounts for a fraction 
of $~10^{-4}$ of the total solar neutrino flux. The important $pp$
flux and the $^7Be$ one which together constitute more than 98\% of 
the total flux have up to now been detected through the inclusive 
measurements of the radiochemical experiments SAGE \cite{Gavrin:2005ks},
\cite{Abdurashitov:2002nt}, Gallex/GNO \cite{Cattadori},\cite{GNO}, 
\cite{Cleveland:1998nv}. Examination of the low energy solar neutrinos in particular
the $pp$ flux alone will teach us about the possible vacuum-matter transition,
test the principle of nuclear energy generation in the sun and the luminosity 
condition \cite{Bahcall:2003ce}. For these reasons performing real time low 
energy solar neutrino experiments should at present be regarded as a major 
objective in the solar neutrino program \cite{Back:2004qi}.

Gallium experiments \cite{Cattadori} are the only ones up to
now in which neutrinos of energy below 1 MeV ($pp,^7\!Be$) account for a 
significant fraction ($\simeq 80\%$) of the event rate. Other experiments are 
unable to detect $pp$ neutrinos, owing to the low threshold required (their 
energies lie below 0.42 MeV) while the $^7Be$ ones account for only $14\%$ 
of the Chlorine event rate \cite{Cleveland:1998nv}. Therefore still 
very little is known about most of the neutrino flux from the sun. Nevertheless, 
as it has been recently noticed \cite{Bahcall:2004ut}, \cite{Cattadori}, 
the average rate from the two Gallium experiments, SAGE and 
Gallex-GNO, has been evolving since the 
time they started in 1990-91 in such a way that the data from the periods 
1991-97 and 1998-03 show a relative discrepancy of $2.4\sigma$ (see table I). 
It is tempting to establish a parallel between this fact and the solar magnetic 
activity. The first period was mainly a time of decreasing activity following
a maximum which had taken place in mid-1990. It ended after the mid-1996 low 
at the initiation of a new solar cycle. For the whole period the average 
sunspot number was 52. In the second
period the solar activity was stronger with a peak in the second quarter 
of 2000 and an average sunspot number of 100 \cite{solar}. While $2.4\sigma$ 
discrepancy is not compelling evidence of new physics, it certainly deserves 
close investigation, especially in view of the above stated fact that Gallium 
are the only experiments with an sizable contribution of $pp,^7\!Be$. 
Consequently, and since no other experiments show such a variational effect, 
the time dependence of these fluxes becomes an open possibility which we 
investigate in the present work. Long-term measurements with low energy solar 
neutrino detectors like the forthcoming Borexino \cite{Borexino}, dedicated 
to $^7\!Be$, and LENS \cite{LENS},\cite{Raghavan:2001jj},\cite{Noon2004} observing 
separately all low energy fluxes, can settle this question.  

The present article aims at exploring and refining a model previously introduced
\cite{Chauhan:2004sf}
based on the joint effect of spin flavour precession to light sterile neutrinos 
and LMA. It will be seen that it can naturally lead to a time dependence of 
the low energy solar neutrino flux (E $<$ 2 MeV) with special incidence on $pp$ 
and $^7\!Be$. To this end the spin flavour 
resonance of these neutrinos must occur in the region where the field is the 
strongest, in the deep convective zone. Their amount of conversion is therefore 
expected to accompany the solar activity. As previously mentioned, the main 
motivation of the present analysis is provided by the Gallium data apparent 
variability and a clear test of the model by the future Borexino and LENS. We 
will therefore present the model predictions for these experiments. 

The article is structured as follows: in section 2 we review the essentials of 
the model, referring the reader to \cite{Chauhan:2004sf} for details. In section
3 we examine Gallium data assumed to be modulated as in table I. We consider 
two options: (a) modulation to be principally due to time dependent $pp$
neutrino conversion and (b) shared between $pp$ and
$^7\!Be$ neutrino conversion. Restricting the oscillation 
parameters $\Delta m^2_{21}, \theta$ within their $1\sigma$ ranges \cite{Gratta}, 
we determine the values of $\Delta m^2_{10}$ (active/sterile mass squared difference), 
$f_B$ ($^8 B$ flux normalization) and field profile which provide the best fits 
separately in each option. All convenient field profiles are expected to exhibit 
a time varying peak in the tachocline correlated with solar activity.  
In the active period (1998-03) the data favour a field profile with an average peak 
value in the range (220-250) kG. For the other, semiquiet period
(1991-97), this decreases to (30-50) kG with a similar profile being favoured.
%within the convective zone at $0.9~R_S$ with an intensity in the range (140-200) kG. 
%The time dependent $^7\!Be$ or intermediate energy conversion [option (a)] 
%is clearly found to lead to the best fits. 
In section 4 we develop the predictions for Borexino and 
LENS assuming the time dependent field profile anchored in the tachocline as derived 
from options (a) and (b). In Borexino the first scenario ($pp$ modulation dominance)
will be more difficult to detect, as expected, while the
second could provide a clear signature. In LENS both cases are visible in each energy 
sector. Finally in section 5 we draw our perspectives and main conclusions, ending 
with a discussion of prospects of active $\rightarrow$ sterile conversion for
supernova dynamics.

\begin{center}
\begin{tabular}{ccc} \\ \hline \hline
Period &  1991-97 & 1998-03 \\ \hline
SAGE+Ga/GNO & $77.8\pm 5.0$ & $63.3\pm 3.6$ \\
Ga/GNO only & $77.5\pm 7.7$ & $62.9\pm 6.0$ \\
no. of suspots     & 52           & 100 \\ \hline
\end{tabular}
\end{center}

{\it{Table I - Average rates for Ga experiments and average number of sunspots 
in the same periods \cite{solar} (units are SNU).}}

\section{Summary of the Model}

The starting point of our present work is a model previously developed based
on LMA with two flavours in which a light sterile neutrino is added
\cite{Chauhan:2004sf}. Its original motivations are the three apparent
problems with LMA: unability to explain the possible time variability of the
neutrino event rate, the predicted upturn of the electron spectrum in
SuperKamiokande (unobserved by experiment) and the 
prediction for the Cl rate (2.9-3.1 SNU) which is about 2$\sigma$ too
high. Decreasing the Cl rate prediction
together with providing a flat spectrum instead of an upturned one
implies a change in the LMA survival probability. The modified
probability should exhibit a dip in the low/intermediate neutrino 
energies. Moreover the conversion from active to sterile state
proceeds through resonant spin flavour precession (RSFP) determined
by a magnetic field profile located mainly nearly the bottom of the 
convective zone of the sun.  The two resonances (LMA and RSFP) therefore
occur at very different solar densities (LMA in the core, RSFP in
the convective zone) and the 'new' mass squared difference between
neutrino flavors is $O(10^{-8} eV^2)$ in order to provide for the
RSFP resonance at the correct location. This choice is not only consistent
with dynamo theories \cite{Antia:2000pu}, which predict a strong field in 
the deep convective zone, but also precludes interference between the 
two resonances, thus providing a clear and observable effect 
superimposed on the 'pure' LMA one. Since, for 
fixed mass squared difference, the neutrino energy determines the location
of the resonance, the time dependent effect associated with a time varying
field profile may affect some of the neutrino fluxes in detriment of others.
The above magnitude of $\Delta m^{2}$ excludes conversion to active neutrinos,
for which both known values of the mass square differences are larger.
So we are lead to consider active $\rightarrow$ sterile neutrino conversion.
Furthermore, conversion of the original $\nu_e$ to an active 
antineutrino \cite{Chauhan:2003wr} (either $\bar\nu_{\mu}$ or $\bar\nu_\tau$) is 
highly disfavoured, since, owing to the large mixing angle, this antineutrino
would oscillate to $\bar\nu_{e}$ on its way to the earth, leading to 
a large observable $\bar\nu_{e}$ flux. This effect, proposed years ago 
\cite{Lim:1990dz},\cite{Akhmedov:1991uk},\cite{Raghavan:1991em},
will not be considered here, as a sizable $\bar\nu_e$ flux is ruled out by 
KamLAND for $E>8MeV$ \cite{Eguchi:2003gg}. There are 
however no low energy limits for $\bar\nu_{e}$ flux from the sun.

In line with our previous work \cite{Chauhan:2004sf}, we will consider
at present the possibility of a time dependent active $\rightarrow$ sterile 
transition. In the simplest such departure 
from the conventional LMA, the active and sterile sectors communicate  
through one magnetic moment transition only, with matter Hamiltonian
\cite{Chauhan:2004sf}

\be
\cal{H}_{\rm {M}}=\left(\begin{array}{ccc}\frac{-\Delta m^2_{10}}{2E}&
\mu_{\nu}B&0 \\ \mu_{\nu}B& \frac{\Delta m^2_{21}}{2E}s^2_{\theta}+V_e&
\frac{\Delta m^2_{21}}{4E}s_{2\theta}\\ 0&\frac{\Delta m^2_{21}}{4E}s_{2\theta}&
\frac{\Delta m^2_{21}}{2E}c^2_{\theta}+V_{x}\end{array}\right)
\ee
in the mass matter basis $(\tilde\nu_0~\tilde\nu_1~\tilde\nu_2)$ to which
corresponds the mixing 
\be
\left(\begin{array}{c}\nu_{s}\\ \nu_{e}\\ \nu_{x}\end{array}\right)=
\left(\begin{array}{ccc}1&0&0\\ 0&
c_{\theta}&s_{\theta}\\ 0&-s_{\theta}&
c_{\theta}\end{array}\right)\left(\begin{array}{c}\nu_{0}\\ \nu_{1}\\
\nu_{2}\end{array}\right).
\ee
in the vacuum basis $(\nu_0~\nu_1~\nu_2)$.
In eqs.(1), (2) $V_e$, $V_{x}$ are the matter induced potentials for 
$\nu_e$ and $\nu_{x}$, $\theta$ is the vacuum mixing angle and $\Delta m^2_{10}=
m^2_{1}-m^2_{0}$ is the mass squared difference between active and sterile 
states.

The important transition whose time dependent efficiency
may determine the possible modulation of neutrino flux is therefore
between mass matter eigenstates $\tilde\nu_0$, $\tilde\nu_1$. This is expected 
to resonate in the region where the magnetic field is the strongest in the
period of high solar activity.

\section{Examining Gallium data}

We refer in this section to Ga data as given in table I, Cl data as
in table II, the SuperKamiokande spectral data for 1496 days as in 
\cite{Fukuda:2002pe} and the SNO data as in \cite{Aharmim:2005gt}. 
Hence we consider time averaged data except for Ga which we split in two
long term sets, namely the averages for 1991-97 (Ga I) and for 1998-03 (Ga II), 
in possible connection to the solar periodic activity. We perform statistical 
analyses for each Ga set together with all other solar data, examining in turn
the case in which the flux modulation is determined mainly by $pp$ neutrinos 
and the case in which the modulation dominance is shared by 
$pp$ and $^7 Be$. These should not however be regarded as two distinct
cases, as they are connected by a continuous evolution of the parameter
$\Delta m^2_{10}$, any intermediate situation being equally viable. 
We consider parameters $\Delta m^2_{21}$ and $\theta$ to be fixed within the 
$1 \sigma$ range of the KamLAND analysis \cite{Eguchi:2003gg}. Hence the 
44 SuperKamiokande spectral data points, 
34 SNO day/night charged current spectral rates, 4 SNO day/night electron scattering
and neutral current rates, the Ga and Cl rates and 2 free parameters ($\Delta 
m^2_{10}$, and the peak field value $B_0$), lead to 82 d.o.f. However, of
these free parameters, the value of $\Delta m^2_{10}$ is fixed from a joint
optimization of 
Ga I and Ga II fits. We evaluate in each case the global $\chi^2$ (rates + spectrum)
referring the reader to \cite{Pulido:2001bd} for definitions. 
Our objective then consists in finding appropriate solar 
field profiles for each of the Ga data sets together with the other solar 
data which provide the best possible fits. The analysis is based on the 
general principle that an intense sunspot activity is correlated with a 
strong field located in the deep convective zone, while in the quiet sun 
period such field may disappear. Throughout the analysis we take $f_B=1.0$, 
the neutrino magnetic moment $\mu_{\nu}=10^{-12}\mu_B$, the LMA mass squared 
difference $\Delta m^2_{21}=8.3\times 10^{-5}eV^2$ and vacuum mixing 
$\theta=0.50$, thus within the KamLAND \cite{Eguchi:2003gg} allowed 1$\sigma$ 
range, and we use the BS05(OP) standard solar model \cite{Bahcall:2004pz}.

\begin{center}
\begin{tabular}{lccc} \\ \hline \hline
Experiment &  Data      &   Theory   &    Reference \\ \hline
Homestake  &  $2.56\pm0.16\pm0.15$ & $8.09\pm^{1.9}_{1.9}$  &
\cite{Cleveland:1998nv} \\
SAGE     &  $see~table~I$ & $125.9\pm ^{12.2}_{12.1}$ & 
  \cite{Abdurashitov:2002nt} \\
Gallex+GNO & $see~table~I$ & $125.9\pm ^{12.2}_{12.1}$ & 
  \cite{GNO}\\
SuperK&$2.35\pm{0.02}\pm{0.08}$ &
$5.69\pm{1.41}$&
\cite{Fukuda:2002pe}\\
SNO CC &$1.68\pm^{0.06}_{0.06}\pm^{0.08}_{0.09}$&$5.69\pm{1.41}$&
\cite{Aharmim:2005gt} \\
SNO ES &$2.35\pm^{0.22}_{0.22}\pm^{0.15}_{0.15}$&$5.69\pm{1.41}$&
\cite{Aharmim:2005gt} \\
SNO NC & $4.94\pm^{0.21}_{0.21}\pm^{0.38}_{0.34}$&$5.69\pm{1.41}$&
\cite{Aharmim:2005gt} \\ \hline
\end{tabular}
\end{center}
{\it{Table II - Data from the solar neutrino experiments except Ga which
is given in Table I. Units are SNU for Homestake and $10^{6}cm^{-2}s^{-1}$
for SuperKamiokande and SNO. We use the BS05(OP) solar standard model 
\cite{Bahcall:2004pz}.}} 

\begin{figure}[h]
\setlength{\unitlength}{1cm}
\begin{center}
%\hspace*{-1.8cm}
\hspace*{-1.6cm}
\epsfig{file=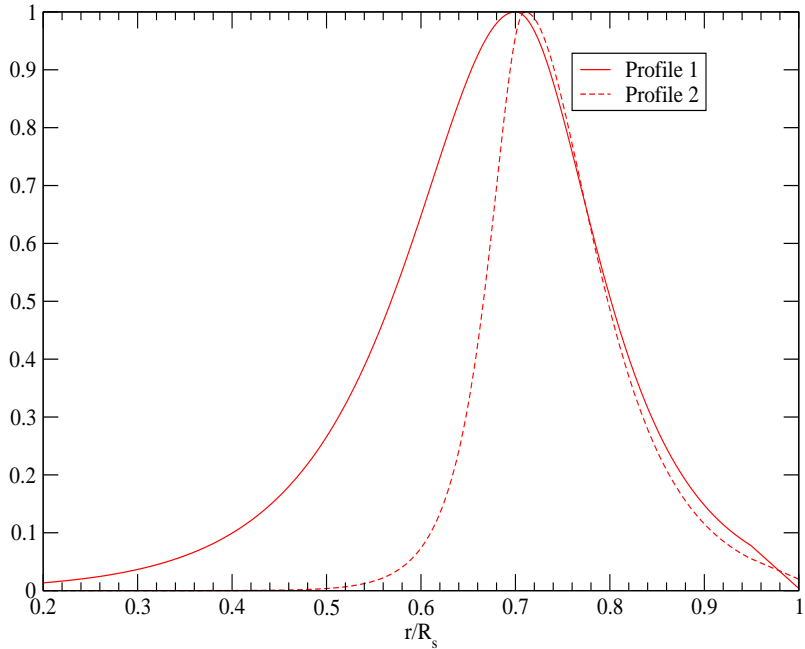,height=13.0cm,width=11.0cm,angle=270}
\end{center}
\caption{ \it Normalized field profiles as a function of the solar coordinate            
$x=r/R_S$. Profile 1: eqs.(3), (4). Profile 2: eqs.(5), (6).} 
\label{fig1}
\end{figure}

\subsection{Modulation by $pp$ }

We start by considering the case where a time varying Ga rate is mainly due
to $pp$ modulation, implying therefore $pp$ resonances to lie in the region of a time
varying field peak. Since this is expected to be located near the bottom of the 
convective zone, thus reflecting the periodic solar activity, this requires 
an active/sterile mass squared difference $\Delta m^2_{10}=O(10^{-8}eV^2)$.
We therefore seek for a set of values of $\Delta m^2_{10}$ and the $^8 B$ flux
normalization factor $f_B$ which provides a good fit for both 
Ga I and Ga II with other solar data, together with a conveniently chosen 
field profile in each situation. The peak field value may be as high as 
$(3-5)\times 10^5~G$ \cite{Antia:2000pu}, \cite{Couvidat:2003ba} at the bottom 
or just below the convective zone in the high activity phase corresponding
to Ga II and much smaller in the semiquiet one (Ga I). 

%We seek for a set of values of $\Delta m^2_{10}$ andflux 
%normalization factor $f_B$ which simultaneously
%provides a good fit for both Ga I and Ga II with other solar data, 
%together with a conveniently chosen field profile in each situation. 
%In this case, for Ga II, the $^7 Be$ neutrino resonance must be located near
%The peak field value which could be as high as $(3-5)\times 10^5~G$
%\cite{Antia:2000pu} \cite{Couvidat:2003ba} at the bottom or just below the
%convective zone and which should nearly vanish for Ga I. Moreover fitting the
%SuperKamiokande spectrum requires reducing its upturn which is characteristic
%of the 'pure' LMA case. While this could be realized through an increase in the 
%vacuum mixing angle $\theta$, such a solution would lead to a strong decrease 
%in the Ga rate, thus preventing the possibility of fitting Ga I: a smaller 
%$\theta$ as compared to its central LMA value is needed if one aims at fitting 
%a Ga rate of 77.8 SNU (see table I), manifestly above its average of 70.8 SNU. 
%An alternative solution to reduce the spectrum upturn is therefore to suppress 
%those $^8B$ neutrinos which mostly contribute to it, i.e. those with energies 
%in the range 5-7 MeV. This implies a second field peak further in the convective 
%zone situated near $0.9~R_S$, provided the $^7 Be$ neutrino resonance lies near the
%first peak at $0.7~R_S$. 
Hence we were lead to the following choice of field profile
for the active phase, Ga II (1998-03) (solid line in fig.1) 
\be
B=\frac{B_0}{ch[10(x-x_{c})]}~~~x_{r}<x<x_{c} 
\ee
\be
B=\frac{B_0}{ch[13(x-x_{c})]}~~~x_{c}<x<x_{r} 
\ee
with $x_r=0.15,~x_c=0.70$ and a peak value $B_0=220~kG$. We take throughout the $pp$
modulation dominance case $\Delta m^2_{10}=-6.0\times 10^{-9}eV^2$ and $f_B=1.0$. With
these choices $pp$ neutrino resonances lie in the range $0.66<x<0.74$ centered near
the peak field value at $x_c$, whereas the main $^7 Be$ line resonance is located at 
$x=0.82$ where the field strength is $B\simeq 0.38B_0$. So the $pp$ modulated case 
also has a non-negligible contribution from $^7 Be$ modulation: otherwise, if the 
time variation were due solely to $pp$ resonances with a negligible field at $^7 Be$ 
ones even in the active period, this would imply an exceedingly fast falling field
in the radial direction, thus worsening the fits.

For the semiquiet phase, Ga I (1991-97), we find the following best choice of field 
profile (dashed line in fig.1)
\be
B=\frac{B_0}{ch[30(x-x_{c})]}~~~x_{r}<x<x_{c} 
\ee
\be
B=\frac{B_0}{ch[15(x-x_{c})]}~~~x_{c}<x<x_{r} 
\ee
with $x_r=0.25,~x_c=0.71$ and $B_0=30~kG$. This is quite similar to the previous one,
the main difference being the peak value. 
The predictions for the 6 rates obtained in the $pp$ modulation dominance in the active 
and semiquiet period are shown in table III. They all lie within $1\sigma$ of 
their central values except for $R_{NC}$ lying at $1.7\sigma$ (see table III).
We note a Ga rate change in a slight excess of $2\sigma$, all other rates being
approximately constant with the possible exception of Cl whose variation is 
nevertheless well within $1\sigma$. In tables III and IV the difference 
$\chi^2_{gl}-(\chi^2_{{SK}_{sp}}+\chi^2_{SNO})$ is the $\chi^2$ 
corresponding to the Ga and Cl rates.

\begin{center}
\begin{tabular}{cccccccccc} \\ \hline \hline
$\rm{B_0(G)}$& Ga & Cl & SK & $\rm{SNO_{NC}}$ & $\rm{SNO_{CC}}$ & $\rm{SNO_{ES}}$ &
$\chi^2_{{SK}_{sp}}$ & $\chi^2_{SNO}$ & $\chi^2_{gl}$/82 d.o.f.\\ \hline
220 kG & 59.6 &  2.67 &  2.26 & 5.66 & 1.56 & 2.23 & 46.4 & 48.9 & 96.4\\
30 kG  & 73.7 &  2.76 &  2.27 & 5.66 & 1.56 & 2.24 & 46.8 & 49.1 & 97.2\\ \hline 
\end{tabular}
\end{center}
{\it{Table III - Peak field values and rates for $pp$ modulation dominance in the 
active period (1998-03) (2nd row) and semiquiet period (1991-97) (3rd row). These 
correspond to field profiles (3), (4) and (5), (6) respectively and $\Delta m^2_{10}
=-6.0\times 10^{-9}eV^2$. $\chi^2_{{SK}_{sp}}$ refers to electron scattering spectrum
and $\chi^2_{SNO}$ to charged current day/night spectrum with in addition the 
4 day/night ES and NC total rates. Units are SNU for Ga, Cl and 
$10^6~cm^{-2}s^{-1}$ for SK and SNO. See tables I, II for a comparison.}}

\begin{figure}[h]
\setlength{\unitlength}{1cm}
\begin{center}
%\hspace*{-1.8cm}
\hspace*{-1.6cm}
\epsfig{file=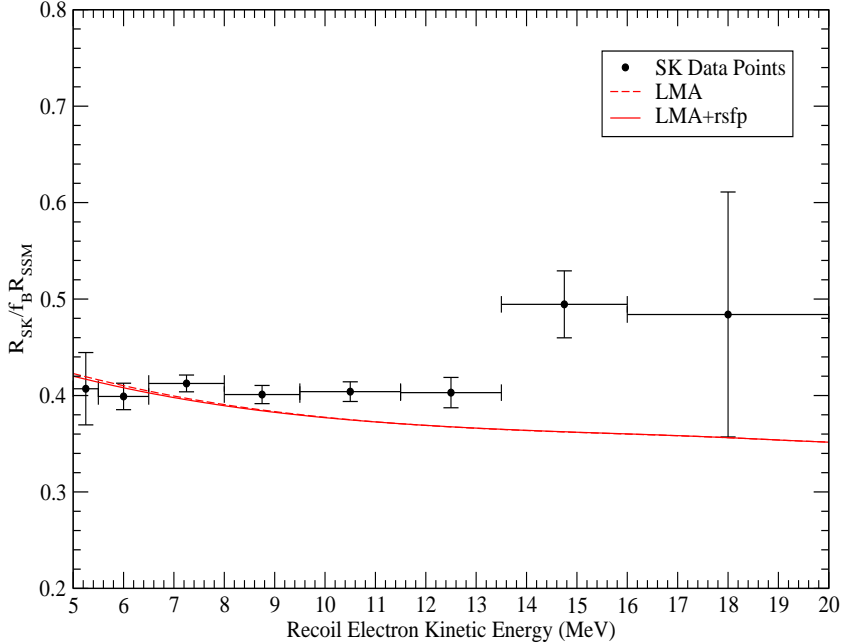,height=13.0cm,width=11.0cm,angle=270}
\end{center}
\caption{ \it SuperKamiokande spectrum normalized to its BS05(OP) standard solar model
\cite{Bahcall:2004pz} value with normalization factor $f_B=1.0$. The typical spectrum 
predicted by the model (full curve) is close to the LMA one (dashed curve).} 
\label{fig2}
\end{figure}

\begin{figure}[h]
\setlength{\unitlength}{1cm}
\begin{center}
%\hspace*{-1.8cm}
\hspace*{-1.6cm}
\epsfig{file=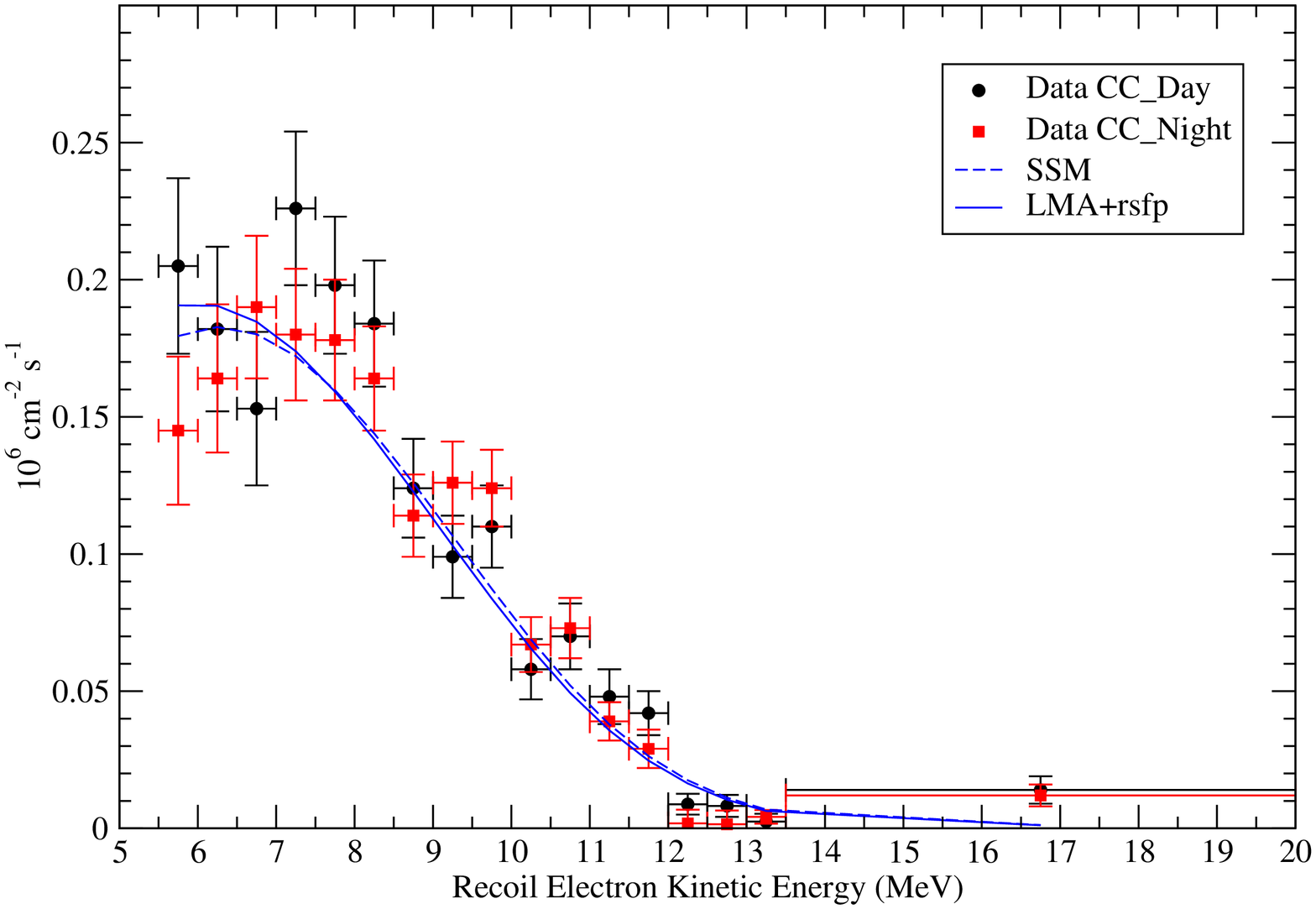,height=13.0cm,width=11.0cm,angle=270}
\end{center}
\caption{ \it SNO charged current spectrum: the model spectrum for all cases (denoted
LMA+RSFP) and the LMA one are practically coincident. SSM denotes the spectrum for
standard neutrinos.} 
\label{fig3}
\end{figure}

\subsection{Modulation by $pp$ and $^7 Be$}

The {\it best} field profiles are in this case the same as the previous ones,
the difference from the former case lying in the parameter $\Delta m^2_{10}$ which 
now satisfies $\Delta m^2_{10}=-1.0\times 10^{-8}eV^2$. All resonances are shifted to 
higher densities with the $pp$ ones located at $0.61<x<0.67$ and the main $^7 Be$ one 
at $x=0.76$. With this choice $^7 Be$ neutrinos have their resonance where the field 
strength is approximately 75\% of its maximum. From table IV, where the rate 
predictions are shown for this case, it is seen that the change in the Cl rate is 
now larger than in the former, owing to the change in $^7 Be$ suppression, being 
however smaller than $1\sigma$. We also note a Ga rate change in excess of $2\sigma$ 
as in the former case. 

Finally, the SuperKamiokande electron scattering
spectrum and the SNO charged current one are shown respectively in figs.2 and 3 
for the active sun: they are practically coincident in the scale of 
figs.2 and 3 for both modulations considered and close to the LMA ones.
 
\begin{center}
\begin{tabular}{cccccccccc} \\ \hline \hline
$\rm{B_0(G)}$& Ga & Cl & SK & $\rm{SNO_{NC}}$ & $\rm{SNO_{CC}}$ & $\rm{SNO_{ES}}$ &
$\chi^2_{{SK}_{sp}}$ & $\chi^2_{SNO}$ & $\chi^2_{gl}$/82 d.o.f. \\ \hline
250 kG & 60.5 &  2.53 &  2.26 & 5.65 & 1.56 & 2.23 & 45.9 & 48.8 & 95.1 \\
50 kG  & 73.6 &  2.75 &  2.27 & 5.67 & 1.57 & 2.24 & 46.5 & 49.1 & 96.9 \\ \hline 
\end{tabular}
\end{center}
{\it{Table IV - Same as table III for the shared $pp$ and $^7 Be$ modulation 
dominance. Here $\Delta m^2_{10}=-1.0\times 10^{-8}eV^2$.}}

\section{Borexino and LENS}

Real time low energy solar neutrino experiments, monitoring $pp$ and $^7 Be$ 
fluxes in a well resolved manner, may test the possible time variability of these 
fluxes as hinted by the Gallium results, thus providing conclusive evidence of the 
neutrino magnetic moment. For this and other important reasons \cite{Bahcall:2003ce}, 
their need was emphasized in the introduction. In this section we present our 
predictions for Borexino \cite{Borexino} and LENS \cite{Raghavan:2001jj},
\cite{Noon2004} experiments. 

\begin{figure}[h]
\setlength{\unitlength}{1cm}
\begin{center}
%\hspace*{-1.8cm}
\hspace*{-1.6cm}
\epsfig{file=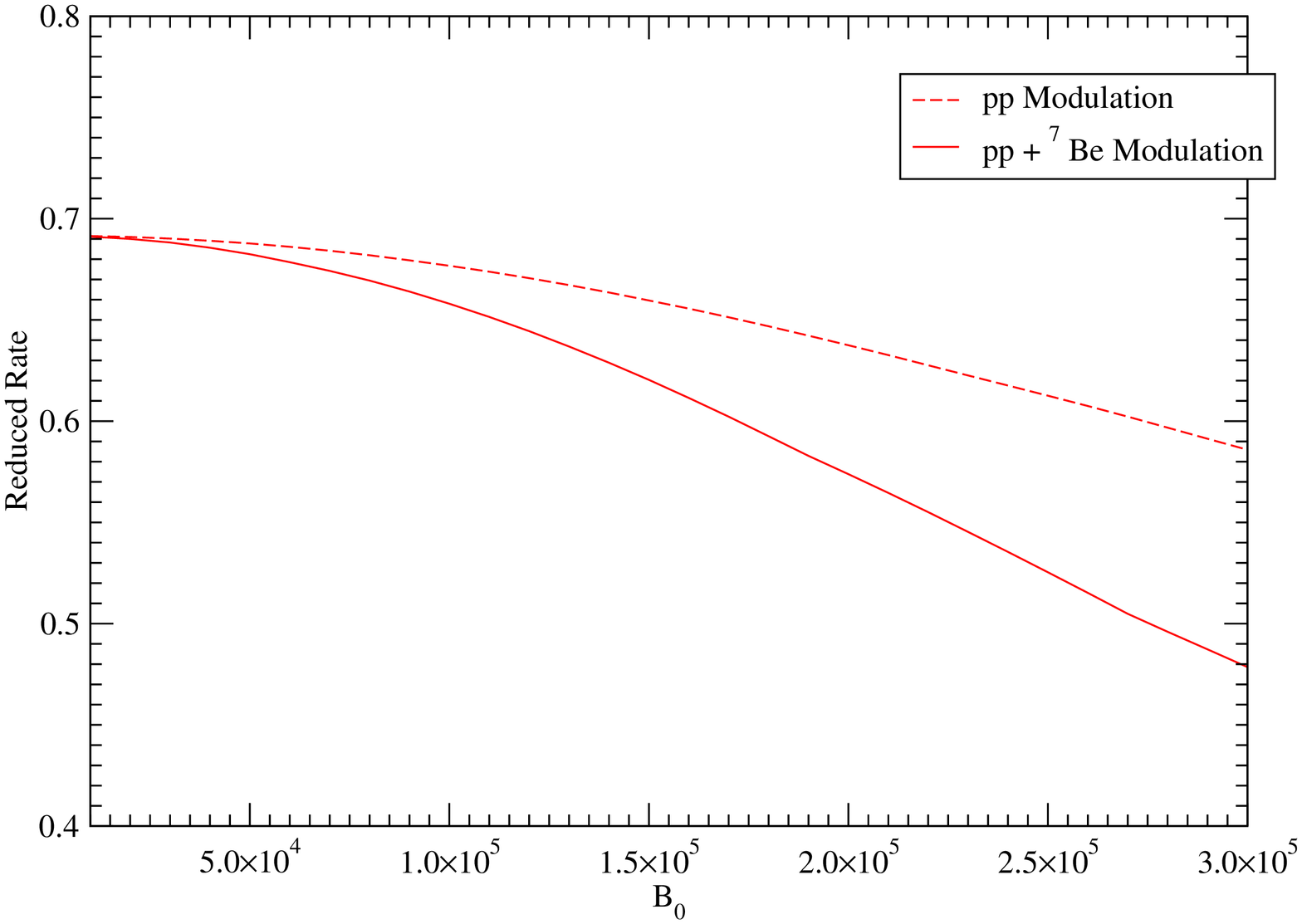,height=13.0cm,width=11.0cm,angle=270}
\end{center}
\caption{ \it Reduced Borexino event rate as a function of the peak field value
(in Gauss) for profile 1.} 
\label{fig4}
\end{figure}

\begin{figure}[h]
\setlength{\unitlength}{1cm}
\begin{center}
%\hspace*{-1.8cm}
\hspace*{-1.6cm}
\epsfig{file=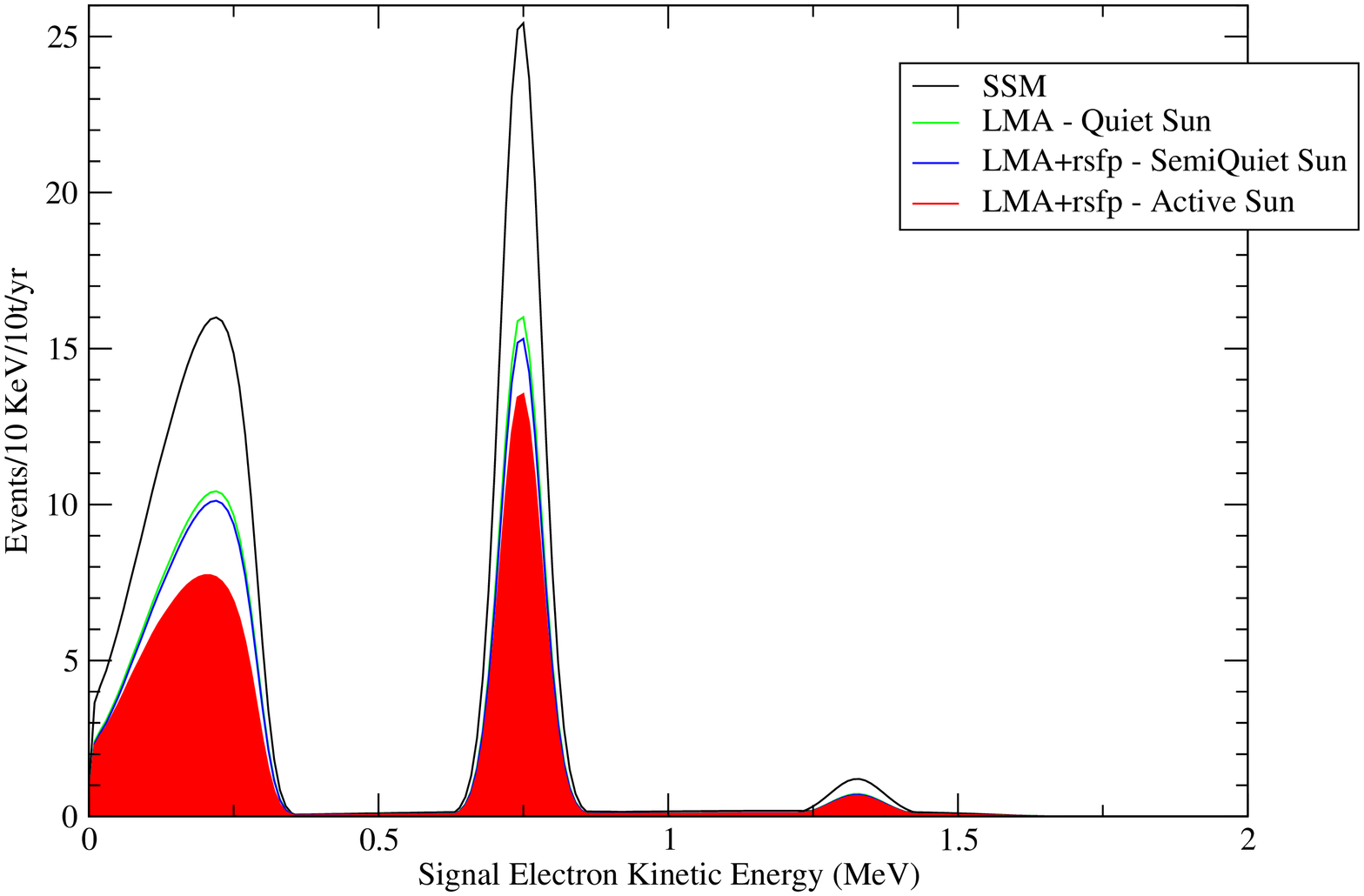,height=13.0cm,width=10.0cm,angle=270}
\end{center}
\caption{ \it Spectral LENS event rates as a function of the measured electron energy.
The upper line (solid) refers to standard neutrinos (no oscillation and no spin flavour
precession). The two middle ones refer to LMA and LMA+RSFP with a peak field value of 50 kG
as considered in the so-called semiquiet case in section 3. The lower line refers to
LMA+RSFP with peak field 220 kG. In both LMA+RSFP cases the $pp$ dominated modulation
is considered (see section 3.1).}
\label{fig5}
\end{figure}

\subsection{Borexino}

Borexino is a real time organic liquid scintillator detector at Gran Sasso aimed at 
measuring the $^7 Be$ flux from the sun. Extremely high radiopurity and very low 
background will allow the detection of record low energy recoil electrons. The 
detection reaction is the neutrino scattering on electrons with a kinetic energy 
threshold of 250 keV and a maximum of 664 keV \cite{Borexino}.
After some technical and environmental problems which caused several year delays, 
water filling is expected to start in the near future and be completed by the
end of 2005. Liquid scintillator filling will then follow, so that
Borexino is expected to start data taking late next year \footnote{For a 
discussion of the general treatment of our Borexino predictions we refer to reader
to \cite{Akhmedov:2002ti}.}. The Borexino collaboration aims at a 10\% total 
statistical and systematic error after one year of run with an improvement to 
5\% after three years.

%\begin{figure}[h]
%\setlength{\unitlength}{1cm}
%\begin{center}
%%\hspace*{-1.8cm}
%\hspace*{-1.6cm}
%\epsfig{file=lens_pp1.ps,height=13.0cm,width=11.0cm,angle=270}
%\end{center}
%\caption{ \it Spectral LENS event rates as a function of the measured electron energy.
%The upper line (solid) refers to standard neutrinos (no oscillation and no spin flavour
%precession). The two middle ones refer to LMA and LMA+RSFP with a peak field value of 50 kG
%as considered in the so-called semiquiet case in section 3. The lower line refers to
%LMA+RSFP with peak field 220 kG. In both LMA+RSFP cases the $pp$ dominated modulation
%is considered (see section 3.1).}
%\label{fig5}
%\end{figure}

We focus our discussion on the dependence of the Borexino event rate on the peak 
field $B_0$ shown in fig.4 for the field profiles considered in section 3, from a 
vanishing field up to a maximum $B_0=300 kG$. We note that for decreasing solar
activity, the requirement of good fits implies a continuous shift in the profile 
(1 $\rightarrow$ 2). 
%The two profiles lead in each case ($pp$ 
%dominated and $pp~+~^7 Be$ dominated modulation) to practically coincident curves on 
%the scale of fig.3 and, for simplicity, we show the curves for profile 1. 
For simplicity in fig.4 we show the curves for profile 1.
We recall that the 'pure' LMA solution corresponds to $B_0=0$, 
so $R_{Bor}=0.69$, as seen from the figure. It is also seen that in the 
$pp~+~^7 Be$ dominated modulation the rate decreases faster for increasing $B_0$,
thus exhibiting more sensitivity to solar activity, than in the $pp$ case.
%a clearer distinction from the 'pure' LMA case 
%is observed than in the $pp$ case. 
In fact for $pp$ dominated modulation the Borexino 
reduced rate varies from 0.69 at $B_0=0$ to 0.59 at $B_0=300~kG$ (0.63 at $B_0=220~kG$), 
while for $pp~+~^7 Be$ it becomes 0.48 at $B_0=300~kG$ (0.53 at $B_0=250~kG$). 
This is to be expected, as Borexino is principally directed at the $^7 Be$ flux:
the more sensitive this flux is to the peak field, correlated to solar activity, the
more sensitive will the Borexino rate be. It is therefore seen that owing to the
size of the experimental errors involved, the active sun (LMA+RSFP) regime may be
clearly distinguishable from the quiet sun (or pure LMA) both for $pp$ and $pp~+~^7 Be$ 
dominated modulation.

\newpage
                                                                                                                            
\begin{figure}[h]
\setlength{\unitlength}{1cm}
\begin{center}
%\hspace*{-1.8cm}
\hspace*{-1.6cm}
\epsfig{file=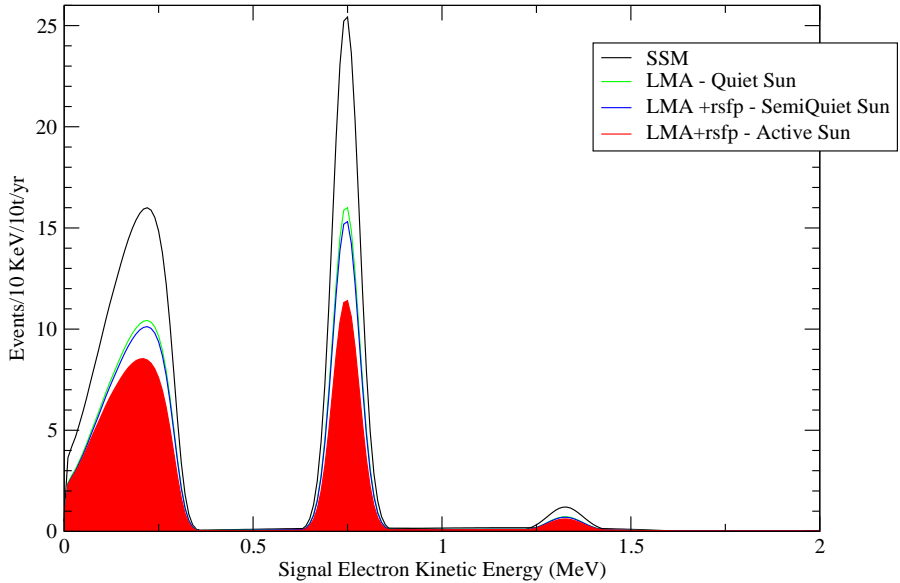,height=13.0cm,width=10.0cm,angle=270}
\end{center}
\caption{ \it The same as fig.5 for the $pp~+~^7 Be$ dominated modulation (see section 3.2)
with a peak field value 250 kG for the lower line.}
\label{fig6}
\end{figure}

%\newpage
\subsection{LENS}

LENS is a real time detector measuring solar neutrinos through the charged
current reaction

\be
\nu_e+^{115}In \rightarrow ^{115} Sn + e^{-}
\ee
with the lowest threshold yet: Q=114 keV \cite{LENS}, \cite{Raghavan:2001jj}. 
Indium was originally proposed in 1976 \cite{LENS} for solar neutrino detection.  
Because of the low threshold, the reaction facilitates access to most of the pp 
continuum. The main technical problem to be solved concerns the background from 
the natural radioactivity of the Indium target itself.
Significant progress in this problem has been made in recent years due to advances  
in the liquid scintillator technology \cite{Noon2004}. Further design innovations 
in 2004 have advanced the project beyond the stage reported in ref. \cite{Noon2004}.

%While this reaction reaches most of 
%the solar $pp$ neutrinos, the idea lay dormant for a long time, the main obstacle being 
%the high background originated from $In$ decay. Recently a new liquid scintillator 
%technology has offered the possibility of reducing this background by a factor of 100 
%or more and a project for a supermodule has been proposed \cite{Noon2004}.

The charged current reaction (7) yields a particularly transparent spectrum, since the 
signal energy is directly and uniquely related to the neutrino energy. A resolved 
spectrum of all low energy components ($pp,~^7 Be,~pep,~CNO$) can be obtained that 
qualitatively shows how the sun shines. 

We have calculated the event rate for the LENS detector in the case of vanishing 
magnetic field ('pure' LMA) and our model profile 1 with LMA for $pp$ 
modulation dominance ($\Delta m^{2}_{10}=-6.0\times 10^{-9}eV^2$) and $pp~+~^7 Be$ 
modulation dominance ($\Delta m^{2}_{10}=-1.0\times 10^{-8}eV^2$). As in Borexino, for 
profiles 1 and 2 the results are practically indistinguishable. The LENS event rate in 
the model is
\be
R_{_{LENS}}=\int_Q^{E_{max}}P_{ee}(E)f(E^{'}_e,E_e)\phi(E)dE.
\ee
Here $E$ is the neutrino energy, $E^{'}_e$ is the prompt (physical) electron energy
($E^{'}_e=E-Q$), $f(E^{'}_e,E_e)$ is the Gaussian energy resolution function with
$\sigma=\frac{\sqrt{NE^{'}_e}}{N}$, $N$ being the signal electron rate/MeV/yr. 
Gaussian resolution functions and detection efficiencies (for optimum
signal/bgd ratios) in current design configurations have been used.
%prompt electron number/MeV.yr in
%a 10kT detector. 
The function $\phi(E)$ represents the standard spectral flux for $pp,~
^7 Be,~CNO,~pep$ neutrinos and $P_{ee}$ is the survival probability. We used detector 
efficiencies $\epsilon =0.35,~0.85,~0.80,~0.90,~0.90$ for $pp,~^7 Be,~N,~O,~pep$ 
neutrinos respectively. LENS event rates are shown in tables V, VI and figs.5, 6.

\begin{center}
\begin{tabular}{cccccc} \\ 
    & $pp$ & $^7 Be$ & $pep$ & $^{13} N $ & $^{15} O $ \\ \hline \hline
Standard & 333.2 &  226.2 &  14.22 & 9.97 & 15.48 \\ \hline
LMA & 211.6 &  138.3 &  8.30 & 6.12 & 9.24 \\ \hline 
LMA+RSFP (semiquiet) & 211.1 & 137.9 & 8.29 & 6.10 & 9.22 \\ \hline 
LMA+RSFP (active) & 171.3 & 120.5 & 7.83 & 5.17 & 8.34 \\ 
\end{tabular}
\end{center}
{\it{Table V - LENS event rates in $pp$ dominated modulation. Units are in 
events/10 t/yr. Parameters are as in section 3.}}

\begin{center}
\begin{tabular}{cccccc} \\ 
    & $pp$ & $^7 Be$ & $pep$ & $^{13} N $ & $^{15} O$ \\ \hline \hline
Standard & 333.2 &  226.2 &  14.22 & 9.97 & 15.48 \\ \hline
LMA & 211.6 &  138.3 &  8.30 & 6.12 & 9.24 \\ \hline 
LMA+RSFP (semiquiet) & 211.4 & 136.2 & 8.26 & 6.04 & 9.16 \\ \hline 
LMA+RSFP (active) & 184.8 & 101.5 & 7.29 & 4.50 & 7.51 \\ 
\end{tabular}
\end{center}
{\it{Table VI - LENS event rates in $pp~+~^7 Be$ dominated modulation. Units are 
in events/10 t/yr. Parameters are as in section 3.}}

\vspace{0.3cm}

Table V, fig. 5 are for $pp$ modulation dominance and table VI, fig.6 for $pp~+~^7 Be$
modulation dominance, all with the parameter values as fixed in section 3. In 
figs.5, 6 the upper curves display the standard neutrino event rates ($P=1$), middle
curves display the 'pure' LMA (quiet sun) and LMA+RSFP event rates in the semiquiet sun 
regime which
are practically coincident as can be seen from the tables. The lower curves are for
the LMA+RSFP rates in the active regime. Here the relatively low value of the $pp$ 
rate is implied by the small detection efficiency ($\epsilon =0.35$) for $pp$ neutrinos, 
and the energy spread seen for $^7 Be$ is originated from the energy resolution function. 
We also note the 0.114 MeV shift toward lower energies of the event rate curve relative
to the solar spectrum.

%\begin{figure}[h]
%\setlength{\unitlength}{1cm}
%\begin{center}
%%\hspace*{-1.8cm}
%\hspace*{-1.6cm}
%\epsfig{file=lens_be1.ps,height=13.0cm,width=11.0cm,angle=270}
%\end{center}
%\caption{ \it The same as fig.5 for the $pp~+~^7 Be$ dominated modulation (see section 3.2)
%with a peak field value 250 kG for the lower line.}   
%\label{fig6}
%\end{figure}

From figs.5 and 6 it is seen that in both cases of study considered in section 3, for
a field of the order of 200 kG in the tachocline the effect of a neutrino endowed with
a magnetic moment is clearly visible in LENS. We recall that the cases considered, which
are defined by the value of the parameter $\Delta m^2_{10}$, span the whole range of
'preferred' fits to the existing data in a model with a field profile which peaks at the 
tachocline. In both cases ($pp$ and $pp+~^7 Be$ modulation dominance) the variation
in the event rate from active sun, assumed to correspond to a tachocline field of 
~200 kG, to semiquiet or quiet (50 kG or less) produces a strong effect in the data and 
is of similar size in both $pp$ and $^7 Be$ sectors.  

\section{Discussion}

In this paper we interpreted the Ga solar neutrino data as providing a hint for long term 
variability of the active solar neutrino flux in possible anticorrelation with sunspot 
activity and attempted at deriving its possible consequences for future experiments, 
namely Borexino and LENS. The claim for such long term variability was first made
for the Cl experiment years ago \cite{Rowley:1984jc},\cite{Voloshin:1986ty},
\cite{Okun:1986hi}, but later turned out to be based on invalid arguments 
\cite{Sturrock:1997gp},
\cite{Walther:1997fy}. The Cl event rate is dominated by high energy neutrinos ($E>5MeV$)
to more than 75\% and the more recent SuperKamiokande experiment, monitoring only these 
high energy ones, did not find any such effect. Long term variability, if it exists, 
is therefore more likely to appear in the low energy sector and its possible observation 
would provide evidence of new physics in connection with the neutrino magnetic moment. 
So far Ga experiments are the only ones having detected the low energy $pp$ and $^7 Be$, 
and they provide some evidence (see table I) of a time varying decay rate which could
be associated to the solar cycle. However $pp$ neutrinos, although overwhelming in
the solar flux, only provide for approximately 55\% of the Ga rate, so their possible
time variation, would be partially 'erased' from the signal, as
they are only seen in an inclusive measurement. The same argument applies to $^7 Be$
neutrinos accounting for 26\% of the rate.

We therefore need real time low energy solar neutrino experiments able to observe 
individually each neutrino component of the spectrum. As the LMA solution 
is based on our incomplete knowledge of the solar neutrino spectrum, one should be 
prepared for surprises in the future. In the previous
sections we listed the main questions left open by the LMA solution (time variability,
too high Cl rate, upturn in the spectrum) and summarized our previous model addressed 
at them using LMA and spin flavour precession to light sterile neutrinos. We attempted 
at fits to data treating separately the 'high' and 'low' Ga rate with a magnetic field 
profile exhibiting a single peak at the bottom or just below the convective zone 
($x=0.7R_S$). We found all rates to be consistent 
with their 1$\sigma$ range except the SNO neutral current one at 1.7$\sigma$. Also our
prediction for the SuperKamiokande spectrum shows the same upturn as the LMA one (see 
fig.2). Concerning this point it should be emphasized that the present sensitivity is 
not enough to make a statistically significant statement. Moreover, decreasing 
the spectrum upturn would require a second field peak at around $0.9~R_S$ which is 
strongly disfavoured. Therefore this aspect should be left open for future clarification
from the SNO experiment. 

We considered time variability associated with the occurence of either the $pp$ or 
$pp~+~^7 Be$ neutrino resonant transition to sterile ones in the region of the strong 
and varying field expected at $x=0.7R_S$. The location of this resonance is fixed by
the active/sterile mass squared difference which must lie in the range $(0.6-1.0)\times 
10^{-8}eV^2$. Our predictions for Borexino and LENS show that these 
experiments have the potential of clearly identifying these solutions at least in the
active solar periods, distinguishing them from the 'pure' LMA ones.

Finally, the proposed mechanism of $\nu_e~\rightarrow~\nu_s$ conversion
is likely to play an important role in supernova dynamics. Its
net result is expected to be the production of a neutron rich
environment, thus facilitating the r-process \cite{McLaughlin:1999pd},
\cite{Fetter:2002xx}. In fact, in the
absence of such conversion, the reaction $\nu_e~n \rightarrow p~e^{-}$,
will play an important role and will lead to the production of alpha
particles via the proton capture of more neutrons. Instead, if
$\nu_e~\rightarrow~\nu_s$ conversion takes place, proton production
is obviously decreased so that more neutrons will be made available
and be rapidly absorbed by seed nuclei, providing an enhancement of
r-process nucleosynthesis.
                                                                                                                            The reduction in the supernova $\nu_e$ flux could probably be clearly
observed in the SNO experiment through the suppression of the charged
current reaction (triggered only by $\nu_e$), while it would be less
apparent in SuperKamiokande where all active neutrinos contribute to
neutrino electron scattering. Furthermore the adiabaticity of the
transition, requiring not only a strong magnetic field $[O(10^{9}~G)]$
but also a smooth density profile, is more likely to be realized in the
later stages of the supernova explosion.

%\newpage
%http://borex.lngs.infn.it/,
%http://www.dxlc.com/solar/solcycle.html   
\vspace{1cm}
\noindent {\Large \bf Acknowledgements}
\vspace{0.5cm}

{\em We acknowledge a discussion with Emanuela Meroni from the Borexino Collaboration.
We are also grateful to David Caldwell for pointing us the importance of
conversion to steriles in the supernova dynamics.
The work of BCC was supported by Funda\c{c}\~{a}o para a 
Ci\^{e}ncia e a Tecnologia through the grant SFRH/BPD/5719/2001.}

%\vspace{0.3cm}  

\end{document}